\documentclass{PoS}

\title{Microquasar interaction with the surrounding medium}

\ShortTitle{$\mu$-quasars/ISM interaction}

\author{\speaker{P. Bordas}
        \thanks{P.B, V. B-R and J.M.P acknowledge support by the Spanish DGPTC of MCI under grant
         AYA2007-6803407171-C03-01, as well as partial support by the European Regional Development Fund
         (ERDF/FEDER). P.B. was supported by the DGPTC of MCI (Spain) under fellowship BES-2005-7234.}\\
         Departament d'Astronomia i Meteorologia and Institut de Ci\`encies del Cosmos (ICC), Universitat de
         Barcelona (UB/IEEC),
         Mart\'{i} i Franqu\`{e}s 1, E08028 Barcelona (Spain)\\
         E-mail: \email{pbordas@am.ub.es}}

\author{V. Bosch-Ramon
       \thanks{V.B-R. gratefully acknowledges support from the Alexander von Humboldt Foundation.}\\
       Max Planck Institut f\"ur Kernphysik, Saupfercheckweg 1, Heidelberg 69117, Germany\\
       E-mail: \email{vbosch@mpi-hd.mpg.de}}

\author{J. M. Paredes$^{\dagger}$\\
       Departament d'Astronomia i Meteorologia and Institut de Ci\`encies del Cosmos (ICC), Universitat de
       Barcelona (UB/IEEC),
       Mart\'{i} i Franqu\`{e}s 1, E08028 Barcelona (Spain)\\
       E-mail: \email{jmparedes@ub.edu}}

\author{M. Perucho
       \thanks{MP acknowledges support from a postdoctoral fellowship of the \emph{Generalitat Valenciana} (\emph{Beca Postdoctoral
       d'Excel$\cdot$l\`encia}), a Max-Planck-Institut postdoctoral fellowship and by the Spanish MCI and the European Fund for Regional
       Development through grants AYA2004-08067-C03-01, AYA2007-67627-C03-01 and AYA2007-67752-C03-02.}\\
       Max-Planck-Institut f\"ur Radioastronomie, Auf dem H\"ugel 69, 53121 Bonn, Germany\\
       E-mail: \email{perucho@mpifr-bonn.mpg.de}}

\abstract{The high kinetic energy outflowing in the jets of microquasars is delivered to the surrounding interstellar medium. This energy input can cause the formation of bow shocks and cocoons that may be detectable from radio to gamma-ray energies. Evidences or hints of emission from jet/medium interactions have been reported for some sources, but little has been done regarding the theoretical modelisation of the resulting non-thermal emission. We have developed an analytical model based on those successfully applied to extragalactic sources for the interaction of AGN jets with their surroundings. Focusing on the adiabatic phase of the growing structures, we give estimations of the expected luminosities through synchrotron, relativistic Bremsstrahlung and inverse Compton processes. We conclude that the interaction structures may be detectable at radio wavelengths, while extreme values for the jet kinetic power, the source age and the medium density are required to make the emission at high and very high energies detectable.\\}

\FullConference{VII Microquasar Workshop: Microquasars and Beyond\\
		 September 1-5 2008\\
		 Foca, Izmir, Turkey}

\begin{document}

\section{Introduction} 

\noindent About 15 microquasars (MQs) are known so far in the Milky Way \cite{Paredes2005}. However, this number could be much higher if other REXBs do in fact display jets but they are to too faint to have been detected \cite{Fender2004a}. MQs show a number of process that resemble those found in extragalactic quasars. For instance, in analogy with radio-loud quasar jets impacting on the intracluster medium, one may expect strong shocks to develope in the termination regions of MQ jets, from where non-thermal radiation should be expected. Hot spots and double-lobe morphologies are common features of the powerful extragalactic FR-II sources \cite{Faranoff1974}. In the case of MQs, hints or evidence of such interactions have been observed in a number of sources (SS~433~\cite{zealey80}, 1E~1740.7$-$2942~\cite{mirabel92}, XTE~J1550$-$564~\cite{corbel02}, Cygnus~X-3~\cite{heindl03}, Cygnus~X-1~\cite{marti96}, \cite{gallo05}, H1743$-$322~\cite{corbel05}, LS~I~+61~303~\cite{paredes07}, Circinus~X-1~\cite{tudose06}). However, only some theoretical work has been done regarding the hydrodynamics of the interaction, or the resulting non-thermal emission in these systems, as compared with the extragalactic case (see, e.g., \cite{aharonian98}, \cite{velazquez00}, \cite{heinz02}, \cite{bosch05}, \cite{Perucho08}).

% 
% \noindent Radio emitting X-ray binaries (REXB) formed by a compact object (a neutron star or a black hole) and a normal (non-degenerate) star, are called Microquasars if they display relativistic jets (Mirabel \& Rodr\'iguez \cite{Mirabel99}). This jets are formed during the so-called Low/Hard state (Fender et al. \cite{Fender2001}), and are expected to be only mildly relativistic (Gallo et al. \cite{Gallo2003}). Presently, about 15 MQs are known in the Milky Way (Paredes \cite{Paredes2005}). However, this number could be much higher if other REXBs do in fact display jets but they are to too faint to have been detected so far (Fender \cite{Fender2004a}).

\noindent From a dynamical point of view, MQ jets can transport large amounts of kinetic energy and momentum far away from the central binary system.
The energy released to the medium can reach $\sim 10^{49}-10^{51}$~erg (\cite{gallo05}, \cite{zealey80}). MQ jet termination regions could produce significant fluxes of non-thermal radiation even for low radiative efficiencies.

% product of the age of the source and the jet kinetic power, $\tau_{\rm MQ}\times Q_{\rm jet}$,
% Additionally, It could well be that MQs were associated, via interaction with their environments, to some of the extended non-thermal radio sources detected in the Galaxy of unknown origin (e.g., Paredes et al. \cite{Paredes07}b).

\noindent We have investigated whether a typical MQ fulfills the conditions to be detectable when interacting with the surrounding external gas. The main paramenters of the model are the MQ kinetic power and age, and the matter densitiy of the environment where the system is embedded in. Furthermore, our model requires certain assumptions regarding the magnetic field value in the interaction regions as well as the acceleration efficiency at the shock fronts. We have fixed them to reasonable values, and we focus our studies on the effects of the dynamical properties of the system and relate them to the final radiation output. We have centered our work in the case of a high-mass system in order to see the role of a massive and hot companion. In the case of a low-mass system, most of the assumptions and conclusions should still be valid, except for the inverse Compton (IC) contribution which would be strongly supressed due to the much fainter radiation field of the companion star.

\noindent In next sections we give an overview of the analytical model used to characterize the shocked emitting zones. The non-thermal fluxes predicted are shown in section Sect.~\ref{res}. We also provide detailed numerical simulations of the hydrodynamical interaction of the jet and the external medium. These simulations do ensure the validity of the assumptions taken in the analytical model, although some differences arise in the description of the recollimation zone (see Sect.~\ref{hydro}). We finally  discuss the obtained results and give some detectability predictions at different energy bands.

\section{Model description}

\noindent The study of the interaction of extragalactic jets with their environments has been extensively adressed from the early 70's up to present days. Pioneer works (\cite{Scheuer1974}, \cite{blandford1974}) stated the main ideas regarding the formation and evolution of the luminous radio lobes present in FR-II sources. We closely follow these works in adapting these models to galactic MQs interacting with the ISM (see also \cite{Kaiser97}) . We consider a supersonic plasma ejected from the central binary system that is decelerated by the external gas accumulated in front of it. This cause the formation of a forward shock extending into the ISM, while a reverse shock propagates inwards into the expelled material. The shocked plasma travelling in the jet inflates a cocoon after crossing the reverse shock. This cocoon is overpressured with respect to the environment although with a much lower mass density, and keeps the jet protected against disruption due to external gas entrainment. In this way, most of the energy and momentum travelling in the jets are delivered at its termination regions without important losses. We concentrate on the adiabatic or Sedov phase \cite{Sedov1959} of the system, in which the jet power is mainly converted in work against the ISM. We take the power and the age of the source, as well as the density of the environment, accordingly to keep the interaction structures in this phase. For simplicity, we consider a constant environment mass density ($\rho_{\rm ISM}$) although possible medium inhomogeneities should be taken into account in a more accurate model. Furthermore, we do not take into account the proper motion of the system. In the case of low-mass systems, the proper motions could play an important role regarding the formation of the shock structures when this velocity is roughly similar to the bow shock velocity \cite{Heinz08}. 

\noindent Fig~1. shows the three different interaction zones where non-thermal radiation can be produced. The first one is the shell region, which corresponds to the material swept up by the bow shock propagating into the ISM. The second one, the cocoon region, corresponds to the shocked material of the jet crossing the reverse shock formed in the collision between the supersonic outflow and the shell (we note that some mixing with the material present in the shocked shell could be expected due to Kelvin-Hemholtz instabilities). The third zone accounts for the recollimation shock formed when the lateral pressure of the jet equals the pressure in the cocoon (see also Sect.~\ref{hydro}).

  \begin{figure}
   \centering 
   \includegraphics[width=9cm]{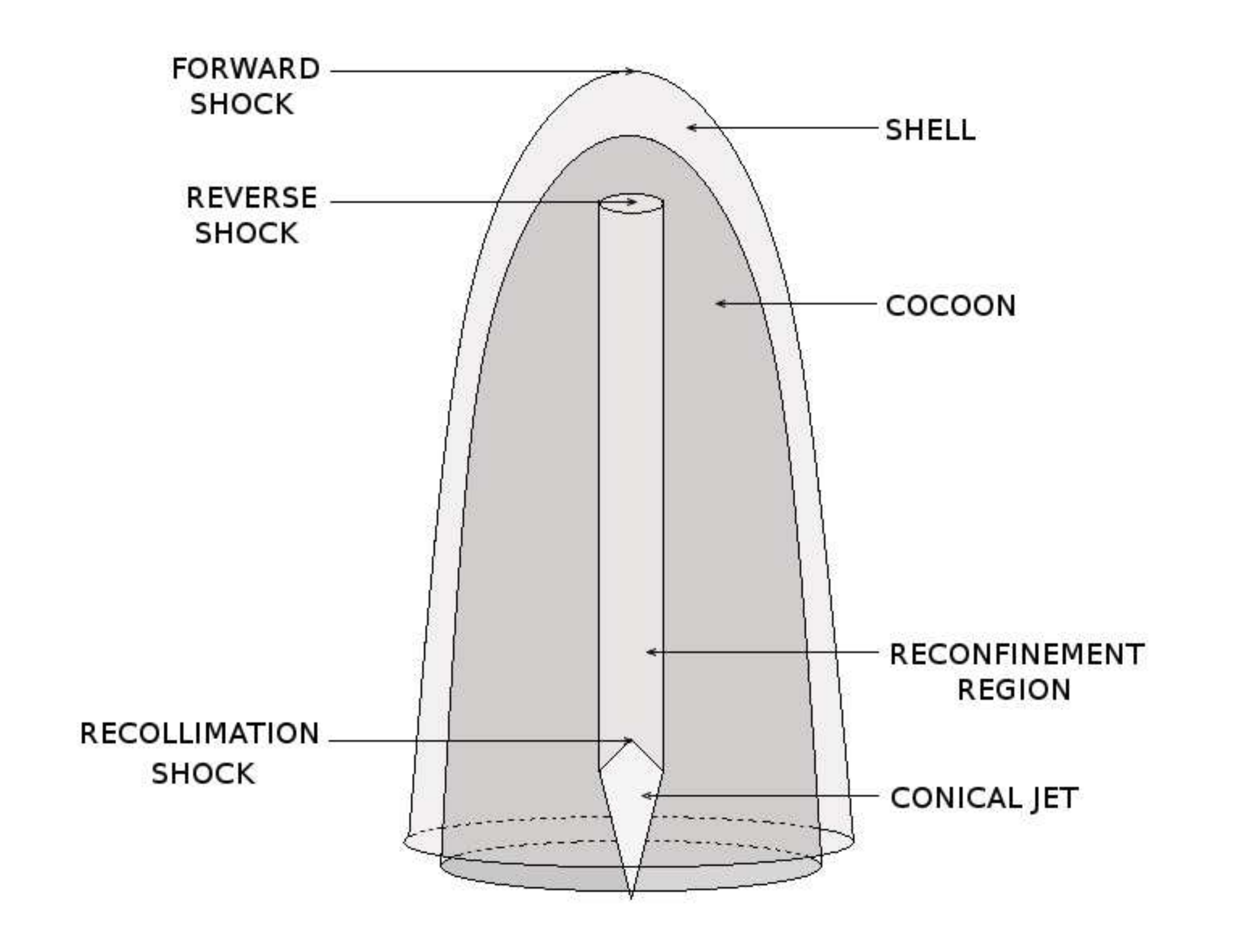}

      \caption{Sketch representing the different interaction zones. The central binary system is located at the bottom, and two jets (here, only one jet is shown) emerge from it and extent outwards until they are effectively decelerated. The jet material that crosses the reverse shock inflates the cocoon, which expands against the shocked ISM. A contact discontinuity separates the cocoon and the shocked ISM, the latter being further limited by the bow shock.}

   \label{fig1}
   \end{figure}

\noindent The conical, ballistic jet emerges from the central engine with an opening angle $\Psi \simeq r/z \sim 0.1$~rad, being $z$ the distance to the injection point and $r$ the jet radius. The jet radius becomes approximately constant at $z_{\rm conf}$, the distance where the lateral jet pressure becomes similar to that of the cocoon, $P_{\rm c}$. The cocoon pressure is balanced with that of the shocked material in the shell, $ P_{\rm b}$, i.e., $P_{\rm c}\approx P_{\rm b}\sim Q_{\rm jet}\times t_{\rm MQ}/V_{\rm b}$, where $V_{\rm b}\sim (4/3)\,\pi\,r_{\rm b}^2 \times l_{\rm b}$ is the bow shock volume, and $r_{\rm b}$ and $l_{\rm b}$ are its radius and length, respectively. Using the Rankine-Hugoniot conditions for strong shocks, the radius of the reconfined jet can be obtained through

\begin{equation}
r_{\rm conf}\sim \Psi \times z_{\rm conf} \sim \Psi \sqrt{\frac{2Q_{\rm jet}v_{\rm jet}}{(\hat{\gamma}+1)(\Gamma_{\rm jet}-1)~\pi c^{2} P_{\rm c}}}\,,
\end{equation}

\noindent where $\hat{\gamma}$ is the adiabatic index of the cocoon material, $v_{\rm jet}$ and $\Gamma_{\rm jet}$ are the jet velocity and bulk Lorentz factor, respectively, and $c$ the speed of light (see, for instance, \cite{Kaiser97}). After crossing the recollimation shock, the jet material still moves at a velocity $\sim v_{\rm jet}$. We neglect further recollimation shocks that may occur (see Sect.~\ref{hydro}).

\noindent The length and the velocity of the bow shock at a given source age, $t_{\rm MQ}$, can be obtained through the equations corresponding to the Sedov phase for the growing structures:
\begin{equation}
l_{\rm b}=c_{1}\left(\frac{Q_{\rm jet}}{\rho_{\rm ISM}}\right)^{1/5}t_{\rm MQ}^{3/5}\,,
\label{aaa}
\end{equation}
where $c_{1}\approx 1$ is a dimensionless constant that depends on the adiabatic index of the jet material and the geometry of the bow shock. The bow shock velocity is:
\begin{equation}
v_{\rm b}=\frac{d}{dt}(l_{\rm b})=\frac{3l_{\rm b}}{5t_{\rm MQ}}\,.
\label{bbb}
\end{equation}

\noindent The radius of the bow shock, $r_{\rm b}$, is found assuming a self-similar relationship between its length and width given by $R\equiv l_{\rm b}/r_{\rm b}$,  We have taken $R=3$ although $R$ can vary from source to source \cite{Perucho07}, \cite{Leahy89}. Given the strong compression of the ISM gas, the length and the width of the cocoon will be also $\sim l_{\rm b}$ and $\sim r_{\rm b}$, respectively. On the other hand, the size of the reconfinement shock is taken as the width of the jet at the location of the recollimation shock, $r_{\rm conf}$, while its size can be up to $\sim l_{\rm b}$. We also take $r_{\rm conf}$ to be rougly the size of the accelerator at the reverse shock, while for the emitter size we take an extension $\sim r_{\rm b}$. In the shell zone, the size of the accelerator and the emitter are taken both to be $\sim r_{\rm b}$. Finally, for the sketch of simplicity, we assume homogeneous conditions in the cocoon and the shell regions, and we apply a one-zone model to compute the non-thermal particle evolution and radiation in these zones. Regarding the magnetic field $B$ in the downstream regions, we derive its value taking the magnetic energy density to be $\sim$~10~\% ($\eta=0.1$) of the thermal energy density.

\begin{table}[!t]
\caption{List of the the parameters which remain with a constant value in the analytic model}             % title of Table
\label{table1}      % is used to refer this table in the text
\centering                          % used for centering table
\begin{tabular}{l c c}        % centered columns (4 columns)

\hline\hline                 % inserts double horizontal lines
{} & {} & {}    \\ 

\it{Parameter} & \it{Symbol} & \it{Value} \\    % table heading 
\hline
\\

Jet Lorentz factor & $\Gamma_{\rm jet}$ & $1.25$\\ [4pt]
Jet half opening angle & $ \Psi $ & 0.1~rad  \\ [4pt]
Luminosity companion star  & $L_{\star}$ & $10^{39}$ erg~s~$^{-1}$ \\ [4pt]
Orbital separation & $ R_{\rm orb} $ &  $3 \times 10^{12}$ cm \\ [4pt]
Self-Similar parameter &  $ R $ & $3$  \\ [4pt]
Magnetic equipartition fraction  & $ \eta $ & $0.1$ \\ [4pt]
Non-thermal luminosity fraction & $ \chi $ & $ 0.01 $ \\ [4pt]

\hline                                   %inserts single line

\end{tabular}
\end{table}

\subsection{Non-thermal particle distribution}

\noindent A power-law particle distribution $N(E)=K\,E^{-p}$, with an spectral index  $p\sim 2$ (see, for instance, \cite{Drury83}), is injected at the reconfinement, reverse and bow shock fronts. The normalization constant $K$ is taken such that $\sim 1$\% of the kinetic power flowing in the jets is converted into non-thermal energy in the postshock regions right after each shock. On the other hand, the maximum energies of the relativistic particles, $E_{\rm max}$, are calculated equating the energy gain to the different cooling processes. The particle energy distribution at $t_{\rm MQ}$ is computed taking into account radiative (synchrotron, relativistic Bremsstrahlung and IC emission; see, e.g., \cite{blum70}) and adiabatic losses. Regarding synchrotron losses, the magnetic field energy density considered above is used. Relativistic Bremsstrahlung is calculated accounting for the densities in the downstream regions. To compute IC losses, we consider only the radiation field from the companion star, an OB star with luminosity $L_{\rm star}=10^{39}$~erg~s$^{-1}$. Adiabatic losses, $\dot{E}=(v/r)\,E$, are computed from the size of the emitters and the expansion velocity. At the downstream region after the reconfinement shock there is no expansion since the jet radius keeps roughly constant. In the case of the shell region, the expansion velocity is $v_{\rm b}$, and for the cocoon, it is $\sim 3/4~v_{\rm b}$ \cite{Landau87}. As noted above, we apply a one-zone model to compute the electron energy distribution in the cocoon and the shell. The electron population properties, the adiabatic coefficient, and the magnetic and radiation fields are homogeneous in both emitting regions. For the electrons injected at the reconfinement shock, their evolution is computed assuming that the stellar radiation density decays as $\propto 1/z^2$. Given $E_{\rm max}$ and the evolved electron distribution in each emitting region, and accounting for the mentioned radiation mechanisms, we can obtain the SEDs for each shocked zone.

\section{Model Results}\label{res}

\begin{figure}[!tp]

   \centering

   \includegraphics[width=15.5cm]{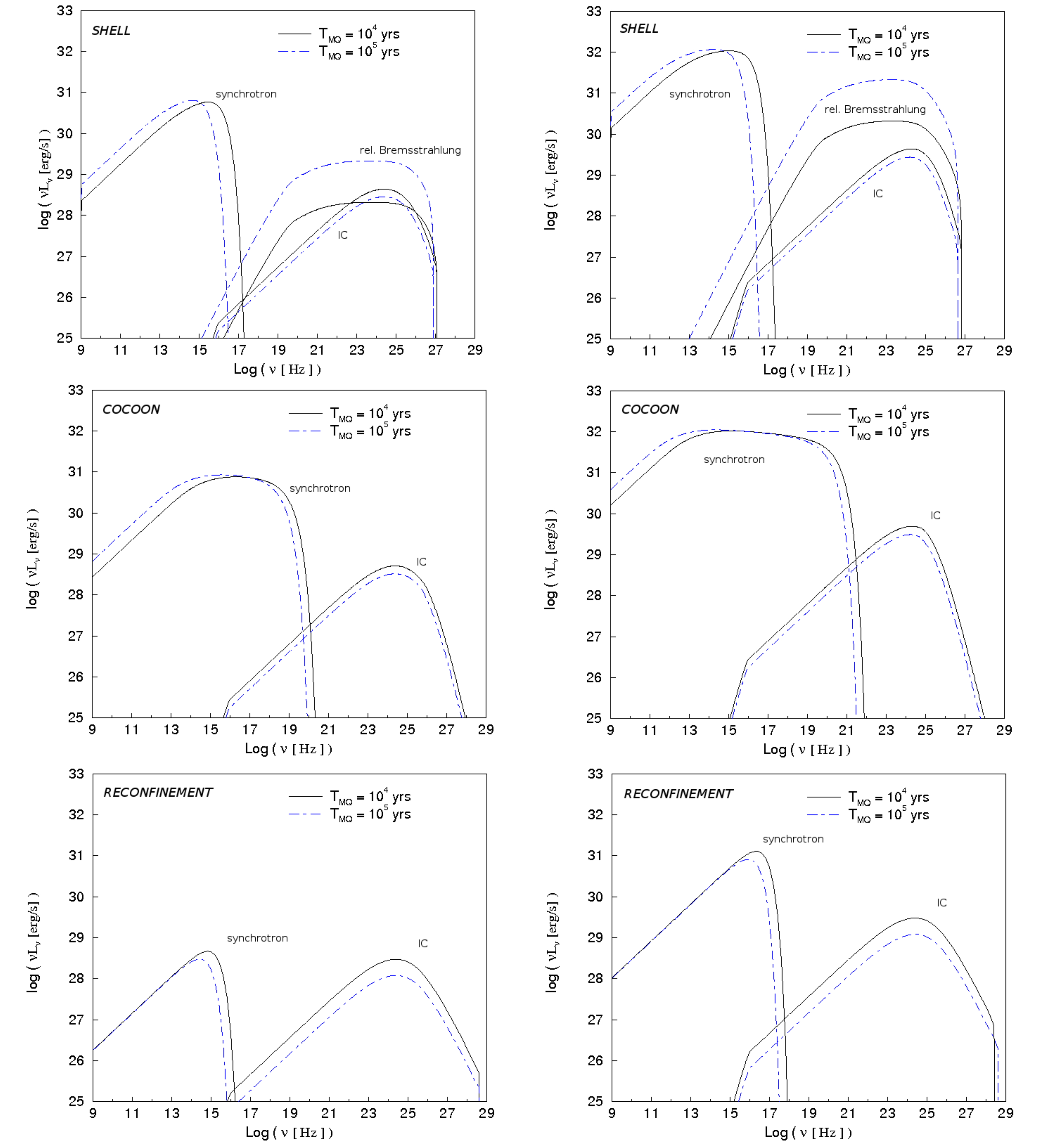} 

   \caption{ Obtained SEDs for the shell (top), cocoon (middle) and jet reconfinement (bottom) regions taking the values for the jet power $Q_{\rm jet}=10^{36}$~erg~s~$^{-1}$ (left panel) and $Q_{\rm jet}=10^{37}$~erg~s~$^{-1}$ (right panel), and an external gas particle density $n_{\rm ISM}=0.1$~cm$^{-3}$ (left panel) and $n_{\rm ISM}=0.1$~cm$^{-3}$ (right panel). Two different values for the source age are represented, $t_{\rm MQ}=$ $10^{4}$~yr (solid lines) and $10^{5}$~yr (dashed lines).}

   \label{fig2}
   \end{figure}

\noindent The effects on the SEDs when varying the source age, $t_{\rm MQ}$, the jet kinetic power, $Q_{\rm jet}$, and the ISM density, $n_{\rm ISM}$ are showed in Fig.~\ref{fig2}. Emission from the shell (top), cocoon (middle) and jet reconfinement (bottom) regions have been computed independently. The contribution of only one jet interacting with the ISM is accounted for in the figures. 

\noindent Maximum energies have been calculated taking into account the accelerator conditions in each zone. In the shell region, $E_{\rm max}$ of electrons range between 2 and 10 TeV, being always limited by synchrotron losses. In the cocoon region, $E_{\rm max}$ can be as high as $\sim$~100~TeV, due to the high velocity of the reverse shock as seen in the jet reference system, and are limited by escape losses, while the same mechanism makes the maximum energies in the reconfinement region to be $E_{\rm max}\sim 3-10$~TeV.

\noindent The maximum emission levels are obtained through synchrotron emission in the three interaction zones, with luminosities that can be as high as $\sim~10^{32}$~erg~s$^{-1}$ for powerful sources (right panel in Fig.~\ref{fig2}). IC emission is the dominant process at the highest energies in the cocoon and reconfinement regions, with luminosities up to $\sim 10^{29}$~erg~s$^{-1}$. Relativistic Bremsstrahlung dominates in the shell in this energy range. The spectrum follows the energy distribution of electrons, and can reach luminosities $\sim 10^{31}$~erg~s$^{-1}$ in the 1~MeV--10~GeV range. This emission channel is on the other hand negligible in the cocoon and reconfinement regions, due the low number density of particles in these zones.

\noindent The source emission response to the variation of $t_{\rm MQ}$ comes mainly through the larger distances from the companion star for older sources and {\it {vice-versa}}. The larger the distance, the fainter the companion star radiation energy density $u_{\star}$ and the lower the IC contribution. Similarly, higher values of $n_{\rm ISM}$ make the jet to be braked at lower distances from the central engine. The interaction regions are then filled with a higher radiation energy density from the companion star, and the IC emission is accordingly enhanced. Finally, all the non-thermal luminosities are proportional to the jet power since the total number of radiation particles depend linearly on $Q_{\rm jet}$.

\noindent Regarding the overall non-thermal emission, the highest radiation output within the set of parameters studied corresponds to the case where $Q_{\rm jet}=10^{37}$~erg~s$^{-1}$, $t_{\rm MQ}=10^{5}$~yr and $n_{\rm ISM}=1$~cm$^{-3}$. Taking a source to be at 3 kpc, a radio flux density of $\sim 150$~mJy at 5~GHz is obtained. Considering an angular extension of a few arcminutes, and taking a radio telescope beam size of $10''$, radio emission at a level of $\sim 1$~mJy~beam$^{-1}$ can be expected. At the X-ray band, we find a bolometric flux in the range 1--10~keV of $F_{\rm 1-10 keV}$ $\sim 2 \times 10^{-13}$~erg~s$^{-1}$~cm$^{-2}$. The X-ray emitting electrons have very short timescales and the emitter size cannot be significantly larger than the accelerator itself. Although the X-rays from the shell are expected to be quite diluted, the X-rays from the cocoon would come from a relatively small region and could be detectable by {\it XMM-Newton} and {\it Chandra} at scales of few arcseconds. In the gamma-ray domain, the flux between 100 MeV and 100 GeV is $F_{\rm 100~MeV<E<100~GeV}\sim 10^{-14}$~erg~s$^{-1}$~cm$^{-2}$, while the integrated flux above 100~GeV is $F_{\rm E>100GeV}\sim 10^{-15}$~erg~s$^{-1}$~cm$^{-2}$. These values are too low to be detectable by current Cherenkov telescopes. For the weakest jets, i.e, lowest ISM densities and youngest sources adopted in our model ($Q_{\rm jet}=10^{36}$~erg~s$^{-1}$, $t_{\rm MQ}=10^{4}$~yr and $n_{\rm ISM}=0.1$~cm$^{-3}$), the fluxes are strongly suppressed. In the radio band, the specific flux is $F_{\rm 5~GHz}\sim 0.1$~mJy~beam$^{-1}$, and the integrated flux at X-rays $F_{\rm 1-10~keV}\sim 10^{-14}$~erg~s$^{-1}$~cm$^{-2}$, and at gamma-rays $F_{\rm 100MeV<E<100GeV}\sim 2\times 10^{-17}$~erg~s$^{-1}$~cm$^{-2}$ and $F_{\rm E>100GeV}\sim$ $2\times 10^{-16}$~erg~s$^{-1}$~cm$^{-2}$.

% 
%   \begin{figure}[!tp]
%    \centering
%    \includegraphics[width=7cm]{TMQ4_5_L36_n1.eps}
%    \caption{Same as in Fig.~2  but taking $n_{\rm ISM}=1$ cm$^{-3}$. See the physical parameter values in Table~2}
%    \label{fig3}
%    \end{figure}
% 
%   \begin{figure}[!tp]
%   \centering
%   \includegraphics[width=7cm]{TMQ4_5_L37_n01.eps}
%   \caption{Same as in Fig.~2  but taking  $Q_{\rm jet}= 10^{37}$ erg~s~$^{-1}$. See the physical parameter values in Table~2}
%   \label{fig4}
%   \end{figure}
% 
%   \begin{figure}[!tp]
%   \centering
%   \includegraphics[width=7cm]{TMQ4_5_L37_n1.eps}
%   \caption{Same as in Fig.~2  but taking  $Q_{\rm jet}= 10^{37}$ erg~s~$^{-1}$ and $n_{\rm ISM}=1$~cm$^{-3}$. See physical parameter values in Table~2}
%   \label{fig5}
%   \end{figure}

\section{Hydrodynamical simulations}\label{hydro}

\noindent In order to check the physical values adopted in the analytical model we have performed numerical simulations of the interaction of MQ with the surrounding medium. A two-dimensional finite-difference code, which solves the equations of relativistic hydrodynamics written in conservation form, has been used (for details, see \cite{Marti97} and \cite{pe+05}). The simulation uses a grid of 320 cells in the radial direction and 2400 cells in the axial direction with physical dimensions of 40$\times\,600$ $r_{\rm j}$. An expanded grid with 320 cells in the transversal direction, brings the boundary from $40\,r_{\rm j}$ to $500\,r_{\rm j}$, whereas an extended grid in the axial direction, consisting of 440 extra cells, spans the grid axially from $600\,r_{\rm j}$ to $900\,r_{\rm j}$. This enlargement ensures that the boundary conditions are sufficiently far from the region of study. The conditions at the boundaries are reflection on the jet axis and in the side where the jet is injected, simulating the presence of the counter-jet cocoon, with the exception of the injection point, where inflow conditions are used. Finally, outflow conditions in the outer axial and radial boundaries are used.

  \begin{figure}[!tp]
  \centering
  \includegraphics[width=15cm]{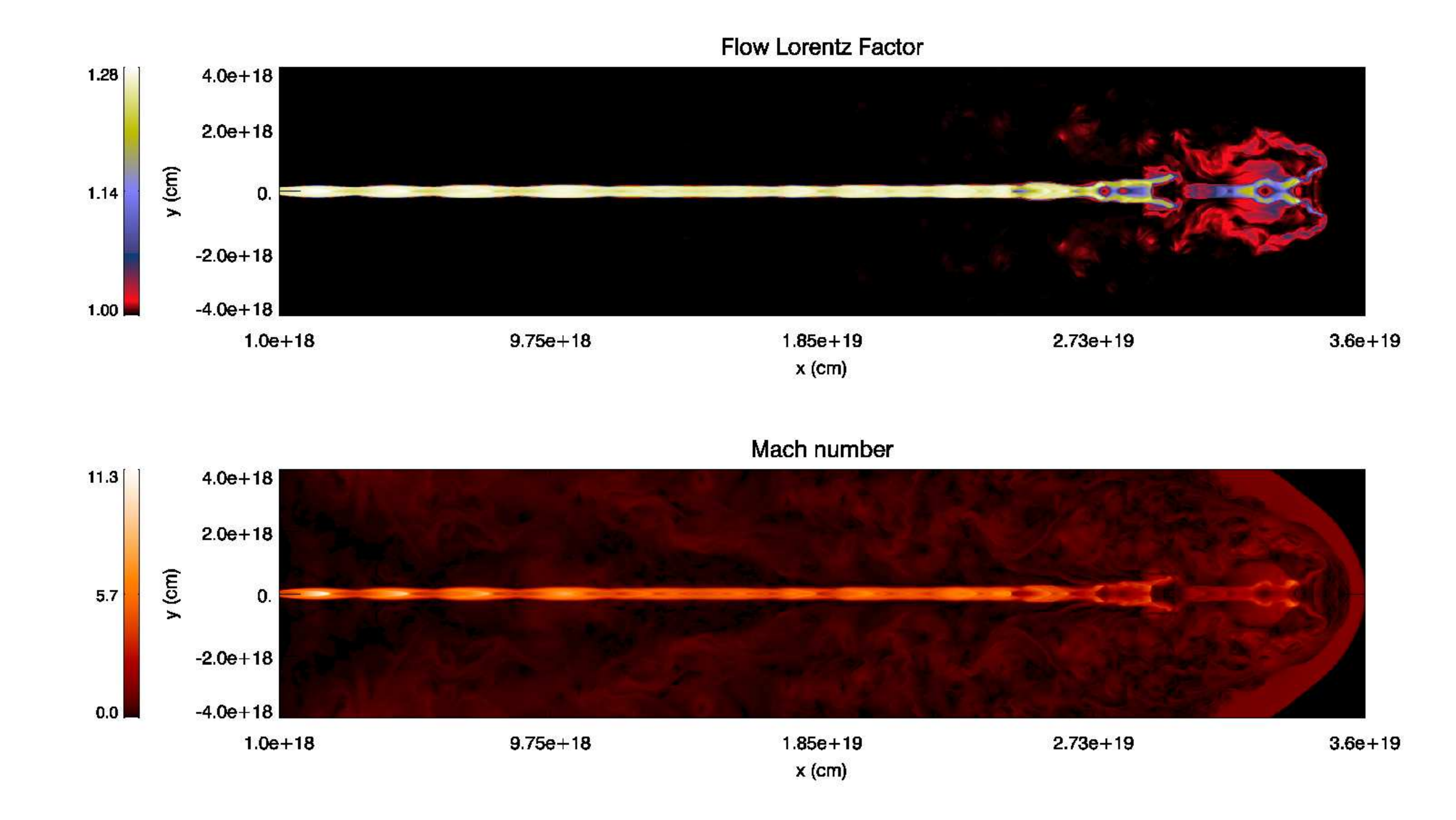}
  \caption{Lorentz factor (top) and Mach number (bottom) maps resulting from hydrodynamical simulations. The simulations were performed using $Q_{\rm
            jet}=3 \times 10^{36}$~erg~s$^{-1}$, $t_{\rm MQ}=3 \times 10^{4}$~yr and $n_{\rm ISM}$=0.3~cm$^{-3}$. }
  \label{vmach}
  \end{figure}

   \begin{figure}[!tp]
   \centering
   \includegraphics[width=15cm]{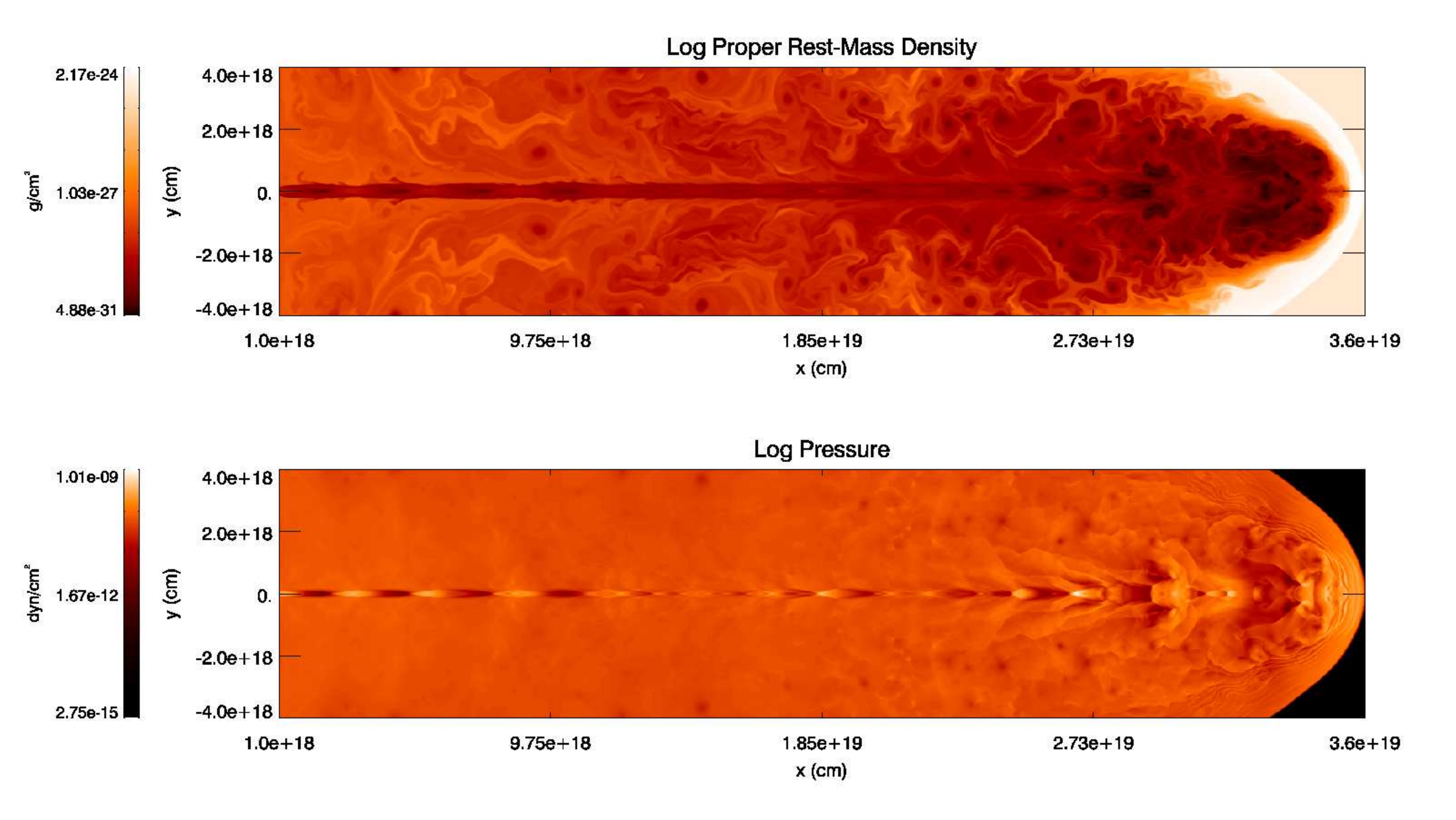}
   \caption{Mass density (top) and pressure (bottom) maps resulting from hydrodynamical simulations. The adopted parameters are the same as those of
            Fig.~3}
   \label{rhopres}
   \end{figure}

\noindent The jet is injected at a distance of $10^{18}$~cm from the compact object, and its initial radius is taken to be $r_0=10^{17}$~cm. The time unit of the code is equivalent to $\approx 3\times 10^6$~s, derived using the radius of the jet at injection and the speed of light ($r_0/c$). Both the jet and the ambient medium are considered to be formed by a non-relativistic gas with adiabatic exponent $\Gamma=5/3$. The number density in the ambient medium is $n_{\rm ISM}=0.3$~cm$^{-3}$. The velocity of the jet at injection is $0.6\,c$, its number density $n_{\rm j}=1.4\times 10^{-5}\,\rm{cm^{-3}}$, and temperature $T\sim 10^{11}$~K (which corresponds to a sound speed $\sim 0.1\,v_{\rm jet}$). These parameter values result in a jet power $Q_{\rm jet}=3\times 10^{36}$~erg~s$^{-1}$. Figs.~\ref{vmach} and \ref{rhopres} show the velocity, Mach number, mass density and pressure maps obtained with the numerical simulations.

\noindent At the time the simulation is stopped, after evolving $\approx 2.7\times 10^4$~yr, the bow-shock is moving at a speed $2-3~\times 10^{7}$~cm~$s^{-1}$, and has reached a distance $\sim 3.6\times 10^{19}$~cm. Initially, the jet expands, accelerating and cooling, due to its initial overpressure. When the flow becomes underpressured with respect to the ambient medium, a first reconfinement shock is generated close to the injection $2\times 10^{18}$~cm. The fluid becomes then again overpressured when passing through the shock and this process is repeated several times around pressure equilibrium with the external medium. At the head of the jet, transonic and subsonic flow velocities result from the increase in temperature and decrease in velocity, as the flow crosses the reverse shock. The cocoon and the shell material are still in high overpressure with respect to the ambient by the end of the simulation. 

\begin{figure*}[!tp]

   \centering

   \includegraphics[width=15cm]{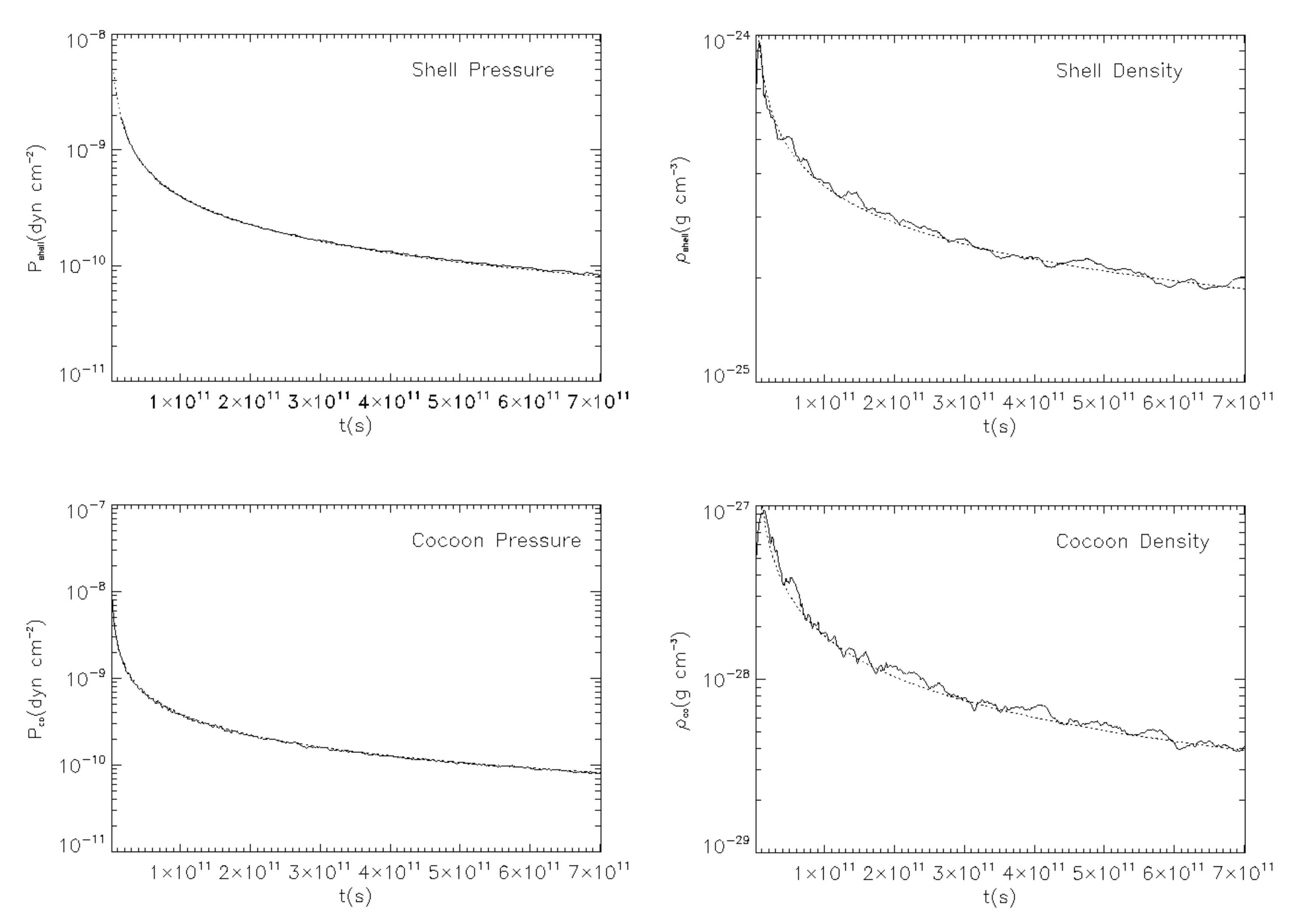}

      \caption{Pressure (left) and mass density (right) evolution in the shell (top) and cocoon (bottom) regions as a function of time. A fitting of the results is also shown for both variables (dotted line). This fit serves to estimate the simulation values at longer times. This extrapolation is strictly valid only if an homogeneous external medium and a constant injection energy rate are assumed. }

         \label{pressions_i_densitats}

   \end{figure*}

\begin{figure}[t]

   \centering

   \includegraphics[width=8cm]{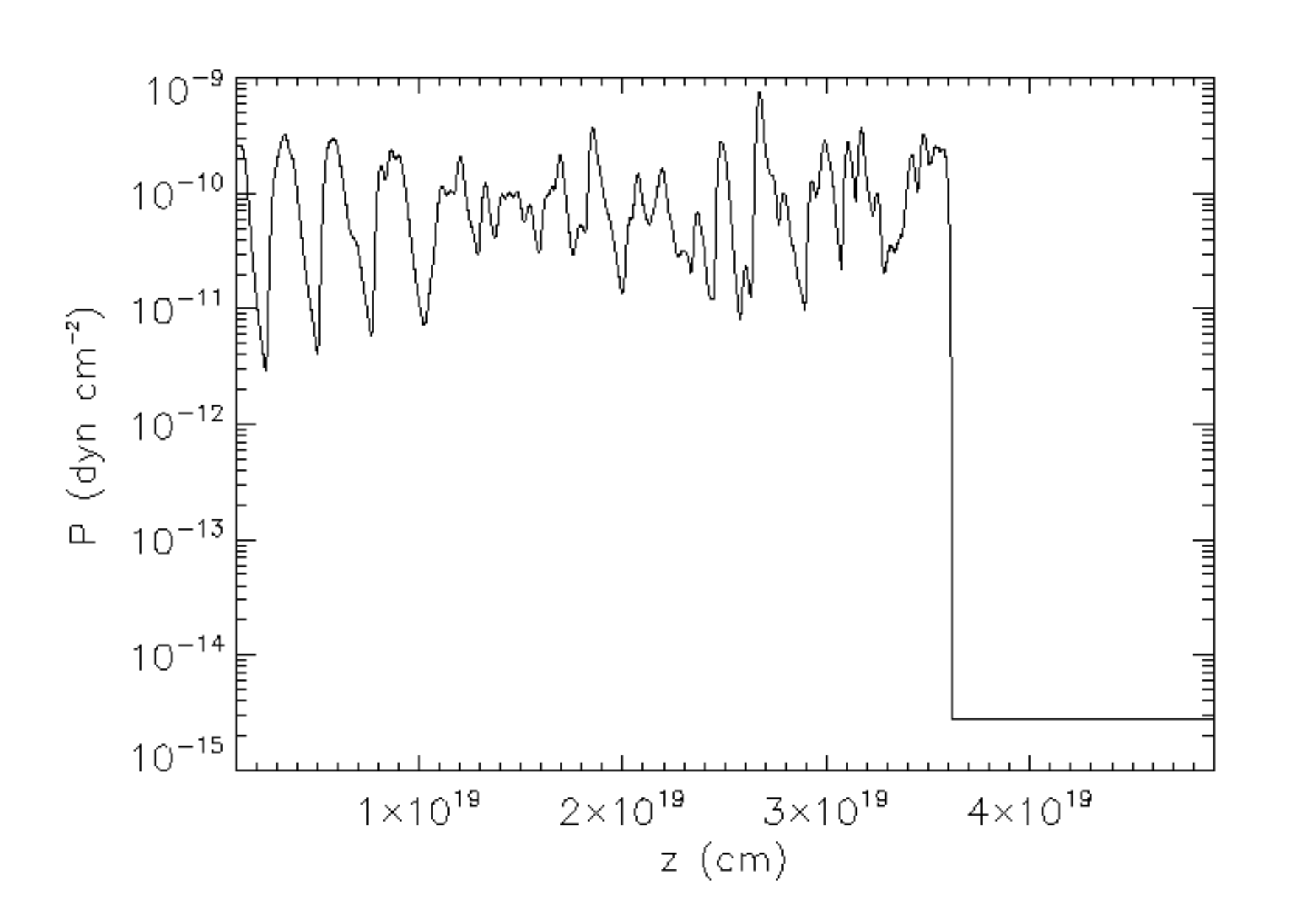}

      \caption{Pressure along the jet axis as a function of distance from its injection point, at $\sim 10^{18}$~cm, as found in the hydrodinamical simulations. Several conical shocks are present, due to the pressure balance with the surrounding cocoon: each time the jet pressure falls below that of the cocoon, a shock is formed, keeping the jet radius roughly constant until it reaches the reverse shock.}

         \label{shocks}
  
 \label{press_axis}
 \end{figure}

\begin{figure*}[!tp]

   \centering

   \includegraphics[width=8cm]{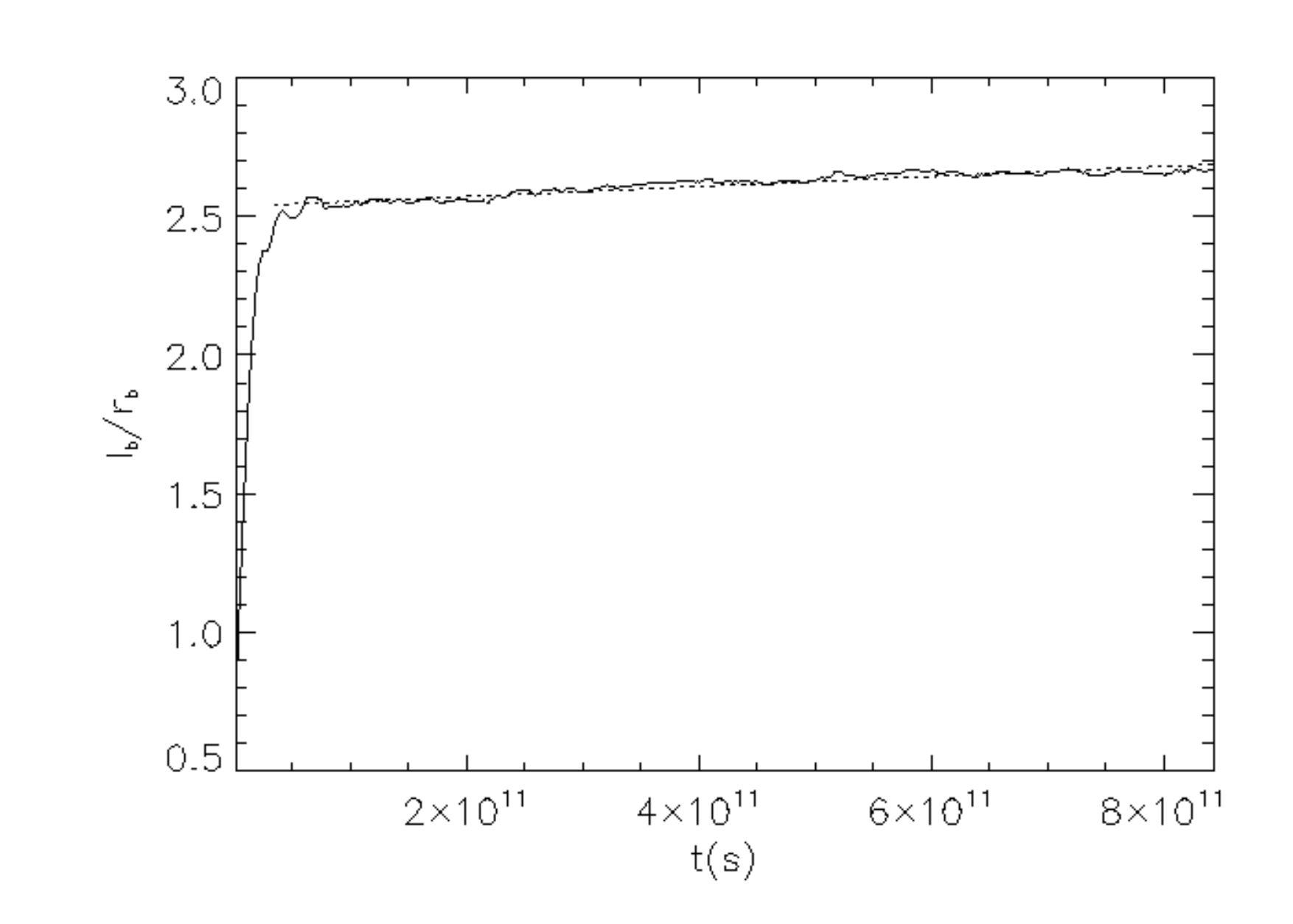}

      \caption{Evolution of the self-similar parameter $R=l_{\rm b}/r_{\rm b}$ as a function of time. After the pronounced initial increase, $R$ remains between 2.5 and 2.7 for most of the simulated time  } 

\label{similar}

   \end{figure*}

\noindent The evolution of the pressure and mass density of the interaction zones in the numerical simulation is shown in Fig.~\ref{pressions_i_densitats} as a function of time, for the shell (top) and cocoon (bottom) regions. The dotted lines in the figures represent the best fits to these evolution plots, which are used in order to extrapolate the values to $t_{\rm MQ} = 10^{5}$~yr. The values obtained are in reasonable agreement with those found in the analytical model. Pressures range between $(2-7)\times 10^{-10}$~erg~cm$^{-3}$, while $\rho_{\rm shell}$ stabilizes at $\sim 2 \times 10^{-25}$~g~cm$^{-3}$ and $\rho_{\rm co} \sim 4 \times 10^{-29}$~g~cm$^{-3}$ for the shell and the cocoon regions, respectively. Moreover, we use a self-similar parameter $R=3$ in the analytical model. We find this value to be in accordance with the results of the numerical simulations, which yield a value between 2.5 and 2.7 as can be seen in Fig.~\ref{similar}. Finally, we note that in our model only a strong shock at the reconfinement point is assumed, while the hydrodynamical simulations show the existence of several conical shocks that develop in the jet when its pressure falls to that of the surrounding cocoon (see Fig.~\ref{shocks}). Therefore, the non-thermal emission presented in Fig.~\ref{fig2} for the reconfinement region should be taken as a rough approximation of the real situation.

\section{Discussion}

\noindent Despite we focus on the non-thermal emission from the MQ jet termination regions, thermal Bremsstrahlung emission should be expected from the shell. Although the shocks considered here are still adiabatic, a non-negligible fraction of the jet kinetic luminosity of up to a few \% may be radiated via thermal Bremsstrahlung. For bow-shock velocities of few times $10^7$~cm~s$^{-1}$, the thermal emission would peak at UV-soft X-rays, energy band that is strongly affected by absorption in the ISM. In general, observations of the thermal radiation from the interaction structures can be still used to extract information of the shell physical conditions (e.g. Cygnus X-1~\cite{gallo05}).
% 
% 
% \noindent The role of accelerated protons in the shocks deserves a few words, since it may be relevant in some specific situations. Given the conditions in the strong shocks we are considering, relativistic protons may reach energies of about 100 TeV; for shell densities $\sim$~1~cm$^{-3}$, the accelerated protons have lifetimes of about 10$^{15}$~s. To reach the gamma-ray fluxes detectable for the present generation of satellite borne and ground based gamma-ray instruments above $\sim 100$~GeV, $\sim 10^{-13}$~erg~cm$^{-2}$~s$^{-1}$, the energy in non-thermal protons stored in the shell should be as high as $\sim 10^{48}$~erg at few kpc distances. Therefore, for a source age of $t_{\rm MQ}=10^{5}$~yr, the required injected power in relativistic protons should be about $\sim 3\times 10^{35}$~erg~s$^{-1}$, thus implying that moderate levels of hadronic emission from MQ jet termination regions may be eventually detected from powerful jets, i.e. $Q_{\rm jet} \gtrsim 10^{37}$~erg~s$^{-1}$.
% 

%Our results, although based on a scenario with a massive and hot star in the central binary system, are applicable as well to a low-mass MQ, but taking into account a much smaller IC cooling. For a low-mass system, the IC component of the predicted SEDs in Sect.~\ref{res} would be several orders of magnitude lower, but the synchrotron and the relativistic Bremsstrahlung components would be roughly the same.

\noindent The reason why some MQs show non-thermal emission from the jet/ISM interaction regions, whereas in other cases such emission remains undetected, is still unclear. In the context of our model, we can study the effects of varying the set of parameters defining the source and their environment, and predict some cases in which the interaction structures may or may not be detectable. First of all, the energy input injected to the medium should be high enough, and the jet kinetic power varies by several orders of magnitude from source to source. In addition, it could be also the case that the density of the surrounding medium is so low that the shell and the cocoon get very large and their radiation too diluted to be detectable \cite{Heinz02}. Moreover, MQ jets could get disrupted at some source age, as it is found in FR~I galaxies. If this happened within times much shorter than the MQ lifetime, the probability to detect a cocoon/shell structure in the MQ surroundings would reduce. On the other hand, some sources may be too far, or the non-thermal fraction too small, to detect significant emission from the interaction regions.

\noindent The evolution of the pressure, mass density, the velocities and the Mach number predicted by the analytical model are in good agreement with those found in the hydrodynamical simulations for the shell and the cocoon regions. Otherwise, several conical shocks may be present within the jet as a consequence of pressure adjustments with the surrounding cocoon, instead of the one strong shock adopted in the analytical treatment. Finally, the length and width of the structures in the model and those found through the numerical simulations are also similar, with a constant ratio $R\sim 3$ in both cases, implying that the physical assumptions used in the analytical treatment are valid to first order.

\noindent The results of this work show that the surroundings of some MQs could be extended non-thermal emitters from radio to gamma-rays, although in the VHE regime the fluxes are too low to be detected. However, taking into account the rough linearity between $Q_{\rm jet}$, $n_{\rm ISM}$, $t_{\rm MQ}$, $\chi$ and $d^{-2}$ with the gamma-ray fluxes obtained, sources with higher values of these quantities than the ones used here (but keeping the adiabatic condition for shocks) may render the MQ jet termination regions detectable by present Cherenkov telescopes. In addition, from a comparison with observations, the magnetic field and the particle acceleration efficiency in the jet/ISM interaction regions can be constrained, giving an insight on the physics of these interaction structures. 

\noindent To conclude, it is interesting to note that, although the adopted model is rather simple, it already accounts for cases when the sources should remain undetectable and cases in which radiation could be detected.


\begin{thebibliography}{99}

% \bibitem{Mirabel99} Mirabel, I. F., \& Rodr\'iguez, L. F. 1999, ARA\&A, 37, 409
% \bibitem{Fender2001}Fender, R. P. 2001, MNRAS, 322, 31
% \bibitem{Gallo2003}Gallo, E., Fender, R.~P., \& Pooley, G.~G. 2003, MNRAS, 344, 60
\bibitem{Paredes2005} Paredes J. M., 2005, in High Energy Gamma-Ray Astronomy: 2nd International
         Symposium, eds. F.~A. Aharonian, H.~J. V\"olk, \& D. Horns. AIP Conference Proceedings, 745, 93
\bibitem{Fender2004a} Fender, R. P. 2004, in Lewin W. H. G., van der Klis M., eds, Compact Stellar X-ray Sources. (Cambridge Univ. Press)
\bibitem{Faranoff1974} Fanaroff, B. L., \& Riley, J. M. 1974, MNRAS, 167P, 31F
\bibitem{zealey80} Zealey, W. J., Dopita, M. A., \& Malin, D. F. 1980, MNRAS, 192, 731
\bibitem{mirabel92} Mirabel, F., Rodriguez, L. F., Cordier, B., Paul, J., \& Lebrun, F. 1992, Nature, 358, 215
\bibitem{corbel02} Corbel, S., \& Fender, R.~P. 2002, ApJ, 573, L35
\bibitem{heindl03} Heindl, W.~A., Tomsick, J.~A., Wijnands, R., \& Smith, D.~M. 2003, ApJ, 588, L97
\bibitem{marti96} Mart\'{\i}, J., Rodriguez, L. F., Mirabel, I. F., \& Paredes, J. M. 1996, A\&A, 306, 449
\bibitem{gallo05} Gallo, E., Fender, R.~P., Kaiser, C., Russell, D., et al. 2005, Nature, 436, 819
\bibitem{corbel05} Corbel, S., Kaaret, P., Fender, R.~P., et al. 2004, ApJ, 617, 1272
\bibitem{paredes07} Paredes, J. M., Rib\'o, M., \& Bosch-Ramon, V., et al. 2007, ApJ, 664, L39
\bibitem{tudose06} Tudose, V., Fender, R. P., \& Kaiser, C. R., et al. 2006, MNRAS, 372, 417
\bibitem{aharonian98} Aharonian, F. A. \& Atoyan, A. M. 1998, NewAR, 42, 579
\bibitem{velazquez00} Vel\'azquez, P. F. \& Raga, A. C. 2000, A\&A, 362, 780  
\bibitem{heinz02} Heinz, S. \& Sunyaev, R. 2002, A\&A, 390, 751
\bibitem{bosch05} Bosch-Ramon, V., Aharonian, F.~A., \& Paredes, J.~M. 2005, A\&A, 432, 609
\bibitem{Perucho08} Perucho, M. \& Bosch-Ramon, V. 2008, A\&A, 482, 917 
% \bibitem{Paredes07} Paredes, J. M., Mart\'{\i}, J., Ishwara Chandra, C. H., \& Bosch-Ramon, V. 2007b, ApJ, 654, L135
\bibitem{Scheuer1974} Scheuer, P. A. G. 1974, MNRAS, 166, 513
\bibitem{blandford1974} Blandford, R. D., \& Rees, M. J. 1974, MNRAS, 169, 395
\bibitem{Kaiser97} Kaiser, C. R., \& Alexander, P. 1997, MNRAS, 286, 215
\bibitem{Sedov1959} Sedov, L. I. 1959, Similarity and Dimensional Methods in Mechanics (New York: Academic Press)
\bibitem{Heinz08} Heinz, S., Grimm, H. J., Sunyaev, R. A., Fender, R. P. 2008, arXiv 0808.1927H
\bibitem{Perucho07} Perucho, M. \& Mart\'i, J. M. 2007, MNRAS 382, 526 
\bibitem{Leahy89} Leahy, J. P., Muxlow, T. W. B., \& Stephens, P. W. 1989, MNRAS, 239, 401
\bibitem{Drury83} Drury, L. 1983, SSRv, 36, 57
\bibitem{blum70} Blumenthal, G. R., \& Gould, R. J. 1970, Rev. Mod. Phys., 42, 237
\bibitem{Landau87} Landau, L. D., \& Lifshitz, E. M. 1987, Fluid Mechanics, 2nd English Edition (Oxford: Pergamon)
\bibitem{pe+05} Perucho, M., Mart\'{\i}, J.~M., \& Hanasz, M. 2005, A\&A, 443, 863
\bibitem{Marti97}Mart\'{i}, J. M. A., Mueller, E., Font, J. A., Ibanez, J. M. A., Marquina, A. 1997, ApJ, 479, 151
\bibitem{Heinz02} Heinz, S. 2002, A\&A, 388, L40


% \bibitem{Falle1991} Falle, S. A. E. G. 1991, MNRAS, 250, 851
%\bibitem[2004]{Fender2004b} Fender, R.~P., Belloni, T.~M. \&, Gallo,~E. 2004, MNRAS, 355, 1105
% \bibitem{Hillas1984} Hillas, A. M. 1984, ARA\&A, 22, 425
% \bibitem{Kaiser04} Kaiser, C. R., Gunn, K. F., Brocksopp, C., \& Sokoloski, J. L. 2004, ApJ, 612, 332
% \bibitem{mart97} Mart\'{\i}, J.M., M\"uller, E., Font, J.A., Ib\'a\~nez, J.M., \& and Marquina, A. 1997, ApJ, 479, 151
% \bibitem{Protheroe1999}Protheroe, R. J. 1999, ADP-AT-98-9 [astro-ph/9812055]
% \bibitem{Ribo2005} Rib\'o, M. 2005, ASPC, 340, 269
%\bibitem[2003]{Wang03} Wang, X. Y., Dai, Z. G., \& Lu, T. 2003, ApJ, 592, 347



% \bibitem{Fanaroff74} Fanaroff B. L., Riley J. M., 1974, MNRAS, 167, 31P
% \bibitem{Heinz08} Heinz, S., Grimm, H. J., Sunyaev, R. A., Fender, R. P. 2008, arXiv 0808.1927H

\end{thebibliography}
\end{document}